\documentclass[aps,prd,10pt,twocolumn,nofootinbib,floatfix,
  noshowpacs,preprintnumbers,superscriptaddress]{revtex4-2}
\usepackage{bm}
\usepackage{epsfig}
\usepackage{amsmath}
\usepackage{tablefootnote}
\usepackage{color}
\usepackage[dvipsnames]{xcolor}

\usepackage{tikz}
\usepackage[compat=1.1.0]{tikz-feynman}
\usepackage{feynmf}

\usepackage[colorlinks=true,breaklinks=true]{hyperref}
\hypersetup{allcolors=[rgb]{0.0 0.0 0.6},linkcolor=teal}
\usepackage[normalem]{ulem}

\newcommand{\exclude}[1]{}

\definecolor{darkolivegreen}{rgb}{0.33, 0.42, 0.18}

\usepackage{fontawesome}
\usepackage{xcolor}
\usepackage{xspace}
\newcommand{\gitlink}{\href{https://github.com/athompson-git/HNL-SBN}{\textsc{g}it\textsc{h}ub~{\large\color{black}\faGithub}}\xspace}

\begin{document}

\title{Enhancing the Sensitivity to Seesaw Predictions in Gauged $B-L$ Scenarios}

\author{Francesco Capozzi}  
\affiliation{Dipartimento di Scienze Fisiche e Chimiche, Università degli Studi dell’Aquila, 67100 L’Aquila, Italy}
\affiliation{Istituto Nazionale di Fisica Nucleare (INFN), Laboratori Nazionali del Gran Sasso, 67100 Assergi (AQ), Italy}

\author{Bhaskar Dutta}
\affiliation{Mitchell Institute for Fundamental Physics and Astronomy, Department of Physics and Astronomy, Texas A\&M University, College Station, TX 77845, USA}

\author{Gajendra Gurung}
\affiliation{Department of Physics, University of Texas, Arlington, TX 76019, USA} 
\affiliation{CERN, Route de Meyrin, 1211 Geneva, Switzerland}

\author{Wooyoung Jang}
\affiliation{Department of Physics, University of Texas, Arlington, TX 76019, USA}  

\author{Ian M. Shoemaker}
\affiliation{Center for Neutrino Physics, Department of Physics, Virginia Tech, Blacksburg, VA 24061, USA}

\author{Adrian Thompson}
\affiliation{Northwestern~University,~Evanston,~IL~60208,~USA}

\author{Jaehoon Yu}
\affiliation{Department of Physics, University of Texas, Arlington, TX 76019, USA}  

\begin{abstract}
New gauge bosons coupled to heavy neutral leptons (HNLs) are simple and well-motivated extensions of the Standard Model. In searches for HNLs in proton fixed-target experiments, we find that a large population of gauge bosons ($Z^\prime$) produced by proton bremsstrahlung may decay to HNLs, leading to a significant improvement in existing bounds on the ($m_{HNL}, U_{\alpha}$), where $U_\alpha$ represent the mixing between HNL and the active neutrinos with flavor $\alpha$.  We study this possibility in fixed target experiments with the 8 GeV proton beams, including SBND, MicroBooNE, and ICARUS, as well as DUNE and DarkQuest at 120 GeV. We find the projected sensitivities to additional $Z^\prime$-mediated HNL production can bring the seesaw mechanism of the neutrino masses within a broadened experimental reach.
\end{abstract}
\preprint{MI-HET-752}
\maketitle

\section{Introduction}
The observation of neutrino flavor oscillations implies the existence of neutrino masses. While a large number of possible scenarios can account for neutrino masses, all of them require physics beyond the Standard Model (SM). A particularly well-studied possibility is the so-called ``seesaw'' scenario in which new right-handed sterile neutrinos mass mix with the left-handed neutrinos in such a way as to explain the smallness of the neutrino masses~\cite{Minkowski:1977sc, Gell-Mann:1979vob, Mohapatra:1979ia}. The existence of such states may be further motivated by new gauge symmetries. In particular, the combination of baryon and lepton number, $B-L$, is non-anomalous in the SM, provided there exist three additional right-handed neutrinos~\cite{PhysRevD.20.776, Mohapatra:1980qe}, making it a favorable extension of the SM.
Thus, gauging the combination of $B-L$ can provide yet additional motivation, beyond neutrino masses, for the necessity of right-handed sterile neutrinos. 

Consider, for example, the right-handed (RH) neutrinos $N_{1,2,3}$, each with a $B-L$ charge of $-1$. Let us simplify the phenomenology of this scenario by only considering interactions with the lightest of the sterile neutrinos, suppose $\nu_4$, which we identify as the HNL $\nu_4 \equiv N$, and denote its mixing matrix element relating it to the active neutrino flavors $\alpha$ as $U_\alpha \equiv U_{\alpha, 4}$. The phenomenological Lagrangian involving a sterile mass eigenstate $N$, for which we take the mixings between $\nu_{R,\alpha}$ and $N$ to be diagonal for simplicity, can then be expressed as
\begin{align}
    \mathcal{L}_{B-L} &\supset 
   - \frac{1}{4} F_{\mu\nu}^\prime F^{\prime \mu\nu} + \frac{1}{2} m_{Z^\prime}^2 Z_\mu^\prime Z^{\prime \mu}   \nonumber \\ 
    &+ g_{B-L} Z^\prime_\mu \sum_{i}  Q_f  \bar{f_i} \gamma^\mu f_i   \nonumber \\
    & -  g_{B-L} Z^\prime_\mu \bar{N} \gamma^\mu N 
\end{align}
where $f = L, e_R, Q, D, u_R, d_R$ for generations $i=1,2,3$. Here $Q_f$ are the $B-L$ charges ($-1$ for the leptons and $+1/3$ for the quarks). Diagonalizing the mass matrix of the neutrinos gives rise to an extended PMNS mixing matrix that includes the elements $U_{e,\mu,\tau}$ between the right-handed and left-handed neutrinos of each flavor. 
Models of gauged $B-L$ such as these that sit below the TeV scale are possible with, for example, $SO(10)$ embedding of $B-L$~\cite{Malinsky:2005bi}, or leptogenesis scenarios~\cite{Abada:2018oly}, and the motivation for a comprehensive search for HNLs down to the $\sim$ MeV mass scale is clear~\cite{Coloma:2015pih, Borzumati:2000mc, Acero:2022wqg}.
In this work, we take a relatively model agnostic set of parameters to search for HNLs in the MeV--GeV mass range, with the relevant physical phenomena described in terms of $g_{B-L}$, $U_\alpha$, $m_{Z^\prime}$, and $m_N$.

Ordinarily, the mixings $U_{\alpha}$ alone can give rise to the production of sterile neutrinos from any process that would also produce active neutrinos. For example, in accelerator targets, one may produce heavy sterile neutrinos from the decays of charged pions, the charged and neutral kaons, and muons, which may take place via the muon and electron mixings, $U_{\mu,i}$ and $U_{e,i}$, respectively. Tau mixings $U_{\tau, i}$, on the other hand, could give rise to sterile neutrino production from tau lepton decays and $D_s$ meson decays~\cite{Ballett:2019bgd}. The sterile neutrino or HNL decay could then proceed via the same mixings in electroweak-mediated decays. 

If the heavy neutrino states are connected to a broken $U(1)_{B-L}$ symmetry, the massive $Z^\prime$ gauge boson could, therefore, contribute additional production mechanisms for the sterile neutrinos if the gauge coupling is not too small. It is this possibility which is the focus of the present study. In this work, we investigate the contribution to enhanced HNL production due to the production of a new vector, $Z^\prime$, which could be produced from proton bremsstrahlung, electron/positron bremsstrahlung, electron/positron annihilation, and neutral meson decays in the proton beam target experiments. We adopt this heuristic model setup of a $Z^\prime$ with gauge coupling $g_{B-L}$ and a single HNL $N$ with mixings to the active neutrinos $U_\alpha$ and calculate the modified model parameter space sensitivity of ongoing and future fixed target experiments; namely, SBND~\cite{SBND}, MicroBooNE~\cite{microboone}, and ICARUS~\cite{icarus} using the booster neutrino beam (BNB), as well as the future DUNE experiment and its near detector (ND)~\cite{Jena:2024ycp}, and the DarkQuest experiment that enjoys a much shorter distance from the beam target to detector~\cite{darkquest}. Similar studies focused on probing gauged $B-L$ with HNLs at the LHC and SHiP~\cite{Batell:2016zod,Deppisch:2019kvs} have indicated that sensitivity to the seesaw parameter space is possible, and in the present work we show that the aforementioned fixed target experiments will also probe the seesaw parameter space through complimentary HNL production channels.

This paper is organized as follows. In \S~\ref{sec:production}, we discuss the production of $B-L$ gauge bosons, $Z^\prime$, in proton beam targets from a variety of production channels that arise from the couplings to SM particles with $B-L$ charge. In \S~\ref{sec:detector}, we discuss the subsequent decays to HNLs via $Z^\prime \to N N$, the propagation of HNLs to the detector and their detection through decays to SM particles. In \S~\ref{sec:results}, we show findings for the sensitivity reach of the ongoing MicroBooNE experiment, as well as future proton beam fixed target experiments SBND, ICARUS, DarkQuest, and DUNE, to HNL decays in the parameter space of their mixing angles and masses for a given $Z^\prime$ gauge boson mass and gauge coupling. Finally, in \S~\ref{sec:conclude}, we summarize our results with concluding remarks on future searches for HNLs.

\section{Production channels of $Z^\prime$}
\label{sec:production}
In proton beam-dump experiments the new $B-L$ gauge boson can be produced either directly via proton bremsstrahlung or indirectly via secondary SM particles. The secondary particles we are interested in are photons and positrons. In the context of $B-L$, each photon can be substituted by a $Z^\prime$ when kinematics allows it. In particular, we consider photons from neutral meson decays and electron bremsstrahlung. On the other hand, positrons can lead to resonant production of $Z^\prime$ through on-shell annihilation with electrons. Here we do not consider $Z^\prime$ production through Compton scattering~\cite{Celentano:2020vtu}, since it has a smaller cross section relative to the other processes and does not change our conclusions in the relevant region of the parameter space. We also neglect charged meson decays, but in \S~\ref{sec:conclude}, we compare our final results with those obtained in ref.~\cite{Berryman:2019dme}, where such a channel is studied in detail.

This Section briefly describes the calculation method we employ for each channel. First, let us define $N_{X}^{ij}$ as the number of particles $X = \pi^0, \eta^0, e^\pm$ in the $i$-th energy bin and $j$-th angular bin as predicted by GEANT4, where the angle is formed by the original proton beam and the outgoing photon propagation directions. We also define the bin extrema to be $[E^{\rm min}_i,E^{\rm max}_i]$ and $[\theta_j^{\rm min},\theta_j^{\rm max}]$ for $i$-th energy and $j$-th angular bins, respectively.

In order to simulate the production of all particles in the proton beam interactions, we used the GEANT4 simulation toolkit~\cite{GEANT4:2002zbu, Allison:2006ve, Allison:2016lfl}. 
The implemented DUNE neutrino production target is a 1.5 m long cylindrical graphite rod with 1.7~cm diameter, following the description in the LBNF beamline design \cite{Papadimitriou:2017zai}. For the BNB experiments, we take a beryllium target cylinder 193 cm long and 1 cm in diameter~\cite{MiniBooNE:2008hfu}. 
We are also interested in HNL production utilizing the BNB beam dump
of dimensions a 4~m wide $\times$ 4~m tall $\times$ 4.21~m long, of which the upstream most 2.64~m long portion stainless steel, followed by a 0.91 m thick concrete and finally a 0.66 m thick stainless steel layers along the beam direction~\cite{MicroBooNE:2015bmn}.

For both the BNB and DUNE target simulations, we used \texttt{QGSP\_BIC\_AllHP} physics list for the hadronic reaction and \texttt{G4EmStandardPhysics} for the electromagnetic interactions. In addition, we have developed an inherited user-defined class of \texttt{G4UserSteppingAction}, derived from \texttt{G4SteppingAction}, to trace and record all the particles produced in the proton beam interactions as they progress throughout the target.  In particular, we recorded 4-momenta of all particles of interest (neutral mesons, electrons, and positrons) produced in the target from the primary proton interaction to the electromagnetic showering process. In order to reduce the computation required, we then bin these particles over their energies and angles with respect to the beam axis, as described previously, with an appropriate binning scheme that is fine enough not to lose important spectral information. The bin weights $N_X^{ij}$ serve as input weights to calculate the $Z^\prime$ production rate, as we describe in the next section.

\subsection{Proton Bremsstrahlung}
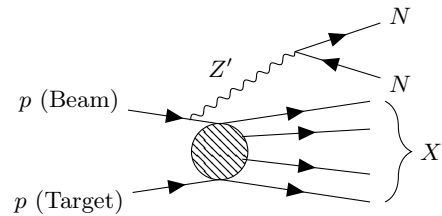
\begin{figure}
    \centering
    \begin{tikzpicture}
    \begin{feynman}
        \vertex (i1) at (-2, 0.67) {$p$ (Beam)};
        \vertex (f1) at (2, 0.67);
        
        \vertex (i2) at (-2, -0.67) {$p$ (Target)};
        \vertex (f2) at (2, -0.67);
        
        \vertex (v1) at (0, 0.37);
        \vertex (v1a) at (-0.4, 0.42);
        \vertex (v2) at (1, 1.33);
        \vertex (v3) at (0, -0.37);

        \vertex (b1) at (0.3, 0.2);
        \vertex (b2) at (0.2, 0.1);
        \vertex (b3) at (0.3, -0.1);

        \vertex (o1) at (2.0, 0.3);
        \vertex (o2) at (2.0, 0.12);
        \vertex (o3) at (2.0, -0.3);
        
        \vertex (f3) at (2.4, 1.8) {$N$};
        \vertex (f4) at (2.4, 0.9) {$N$};

        \vertex [blob] (a) at (0,0) {};

        \draw [decorate,decoration={brace,amplitude=10pt}] (2.2,0.67) -- (2.2,-0.67) node[midway,xshift=10pt,right] {$X$};

        \diagram* {
            (i1) -- [fermion] (v1) -- [fermion] (f1),
            (i2) -- [fermion] (v3) -- [fermion] (f2),

            (b1) -- [fermion] (o1),
            (b3) -- [fermion] (o3),
            
            (v1a) -- [boson, edge label=$Z^\prime$] (v2),

            
            (v2) -- [fermion] (f3),
            (v2) -- [anti fermion] (f4),
        };
    \end{feynman}
\end{tikzpicture}
    \caption{Proton bremsstrahlung via non-single diffractive scatttering, producing the on-shell $U(1)_{B-L}$ gauge boson that promptly decays to $NN$ pairs.}
    \label{fig:pbrem}
\end{figure}
First, we consider the $Z^\prime$ bremsstrahlung of the primary beam protons, Fig.~\ref{fig:pbrem}. The diffractive scattering process is modeled in the quasi-real approximation by considering producing the vector $Z^\prime$ as initial state radiation (see ref.~\cite{Foroughi-Abari:2021zbm} for details). We briefly summarize the calculation as follows. The differential scattering cross-section is given by
\begin{equation}
\label{eq:pbrem}
    \dfrac{\partial^2\sigma(p p \to X Z^\prime)}{\partial p_T^2 \partial z} = w^D(z,p_T^2) \times \sigma_{pp}^\text{NSD}(s^\prime)
\end{equation}
where $w^D(z,p_T^2)$ is a splitting function over the longitudinal momentum fraction $z = p_{Z^\prime}/p$ of the outgoing $Z^\prime$ momentum with respect to the initial proton momentum $p$, $p_T$ is the transverse momentum of the outgoing $Z^\prime$, and $s^\prime = 2m_p (p(1-z)+m_p)$ where $m_p$ is the proton mass. In the construction of the splitting function $w^D(z,p_T^2)$ in ref.~\cite{Foroughi-Abari:2021zbm}, there are hadronic form factors introduced to capture timelike momentum transfer as well as the off-shell momentum in Fig.~\ref{fig:pbrem}. The parametrization of the form factors can have a significant impact on the magnitude of the cross section (up to an order of magnitude), especially in the parameter $\Lambda_p \sim \mathcal{O}(m_p)$ which acts as a cutoff scale for the off-shell behavior. In this work we take the central value in the range considered by ref.~\cite{Foroughi-Abari:2021zbm} for the cutoff scale, $\Lambda_p = 1.5$ GeV. 

In the integration of Eq.~\ref{eq:pbrem}, certain cuts have to be made to ensure the applicability of the quasi-real approximation; following ref.~\cite{Foroughi-Abari:2021zbm}, we define
\begin{align}
    \Theta_{cuts} &\equiv \Theta\bigg( 0.2 - \frac{H(z,p_T^2)}{4 z(1-z)^2 p^2} \bigg) \Theta (0.2 - p_T/E_p) \nonumber \\
    &\times \Theta(0.2 - m_{Z^\prime}/E_{Z^\prime})
\end{align}
where $H(z,p_T^2) \equiv p_T^2 + z^2 m_p^2 + (1-z)m_{Z^\prime}^2$ is a kinematic structure function.

Assuming that most of the protons get absorbed in the proton target, we can estimate the differential rate of $Z^\prime$ particles produced by proton bremsstrahlung as the number of protons on target (POT) times the ratio of the bremsstrahlung and total proton cross sections~\cite{Berryman:2019dme, Dev:2023zts};\footnote{During the completion of this work, other analyses have proposed updated treatments of the form factors used proton bremsstrahlung, see for example ref.~\cite{Foroughi-Abari:2024xlj} and refs.~\cite{Gorbunov:2024vrc, Gorbunov:2023jnx}. The potential impact of these new calculations is left to a future work.}
\begin{equation}
     \dfrac{\partial^2 N_{Z^\prime}}{\partial p_T^2 \partial z} = N_\text{POT} \times \dfrac{1}{\sigma_\text{tot}(s)} \times  \dfrac{\partial^2\sigma(p p \to X Z^\prime)}{\partial p_T^2 \partial z} \times \Theta_{cuts}
\end{equation}

\subsection{Neutral Meson Decay} 
\begin{figure}
    \centering
    \begin{tikzpicture}
        \begin{feynman}
         \vertex (o1);
         \vertex [left=1.0cm of o1] (i1) {$\pi^0, \eta$};
         \vertex [above right=1.2cm of o1] (i2) {$\gamma$};
         \vertex [below right=1.2cm of o1] (f1);
         \vertex [above right=1.0cm of f1] (ff1) {$N$};
         \vertex [below right=1.0cm of f1] (ff2) {$N$};
         \diagram* {
           (f1) -- [boson, edge label=$Z^\prime$] (o1) -- [boson] (i2),
           (o1) -- [scalar] (i1),
           (ff1) -- [fermion] (f1) -- [fermion] (ff2)
         };
        \end{feynman}
       \end{tikzpicture}
    \caption{Neutral meson decay to a photon and a $B-L$ gauge boson $Z^\prime$, and the subsequent on-shell decay to HNL pairs.}
    \label{fig:meson_decay}
\end{figure}
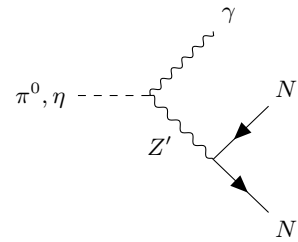

The number of $Z^\prime$ from $N_M^{i,j}$ mesons in the $(i,j)$-th (energy, angle) bin of the neutral meson GEANT4 monte carlo sample can be estimated by
 \begin{equation}
 N_{Z^\prime}^{M,ij}=N_{M}^{ij}\frac{{\rm Br}(M\to\gamma Z^\prime)}{2{\rm Br}(M\to\gamma\gamma)}\,,
 \label{N_meson_bin}
 \end{equation}
where $M=\pi^0,\eta^0$ and ${\rm Br}(M\to\gamma Z^\prime)$ is the branching ratio of the neutral meson decaying into a photon and a $Z^\prime$, shown in Fig.~\ref{fig:meson_decay}. The branching ratio can be expressed as
 \begin{equation}
 {\rm Br}(M\to\gamma Z^\prime)=2\left(\frac{g_{B-L}}{e}\right)^2\left(1-\frac{m^2_{Z^\prime}}{m^2_{M}}\right)^3 {\rm Br}(M\to\gamma \gamma)\,,
 \label{branching_ratio_meson}
 \end{equation}
where  ${\rm Br}(\pi^0\to\gamma \gamma)=0.98823$ and ${\rm Br}(\eta^0\to\gamma \gamma)=0.3941$ \cite{Zyla:2020zbs}. For each meson in the GEANT4 sample, we perform a 2-body decay monte carlo for $M \to \gamma Z^\prime$ in the rest frame, then boost to the lab frame to simulate the boosted spectrum of $Z^\prime$. 

\subsection{Electron and Positron Bremsstrahlung}
\label{sec:ebrem}

\begin{figure}
    \centering
    \begin{tikzpicture}
              \begin{feynman}
         \vertex (o1);
         \vertex [left=1.1cm of o1] (i1) {$e^\pm$};
         \vertex [right=1.1cm of o1] (f1) {$e^\pm$};
         \vertex [below=1.1cm of o1] (o2);
         \vertex [right=1.1cm of o2] (f2) {$A$};
         \vertex [left=1.1cm of o2] (i2) {$A$};
         \vertex [left=0.7cm of o1] (b1);
         \vertex [above right=1cm of b1] (f3);
         \vertex [above right=0.7cm of f3] (n1) {$N$};
         \vertex [right=0.8cm of f3] (n2) {$N$};

         \diagram* {
           (i1) -- [fermion] (o1) -- [fermion] (f1),
           (i2) -- [fermion] (o2) -- [fermion] (f2),
           (o2) -- [boson] (o1),
           (b1) -- [boson, edge label=$Z^\prime$] (f3),
           (n1) -- [fermion] (f3) -- [fermion] (n2)
         };
        \end{feynman}
       \end{tikzpicture}
    \begin{tikzpicture}
        \begin{feynman}
         \vertex (o1);
         \vertex [above left=1cm of o1] (i1) {$e^-$};
         \vertex [below left=1cm of o1] (i2) {$e^+$};
         \vertex [right=1.2cm of o1] (f1);
         \vertex [above right=1cm of f1] (ff1) {$N$};
         \vertex [below right=1cm of f1] (ff2) {$N$};
         \diagram* {
           (i1) -- [fermion] (o1) -- [fermion] (i2),
           (o1) -- [boson, edge label=$Z^\prime$] (f1),
           (ff1) -- [fermion] (f1) -- [fermion] (ff2),
         };
        \end{feynman}
       \end{tikzpicture}
    \caption{Left: Bremsstrahlung of $Z'$ off of secondary electron and positron in their interactions with the target atoms,  $A = ^{12}$C (DUNE target), $A=^9$Be (BNB target) or $A=^{56}$Fe (DarkQuest), that radiate $B-L$ $Z^\prime$ and subsequently decay to HNL$NN$ pairs. A similar diagram where the $Z^\prime$ is radiated off the $e^\pm$ final state also contributes. Right: positron annihilation on target electrons to resonantly produce an on-shell $Z^\prime$ and its subsequent decay to $NN$ pairs (right).}
    \label{fig:brem-res}
\end{figure}
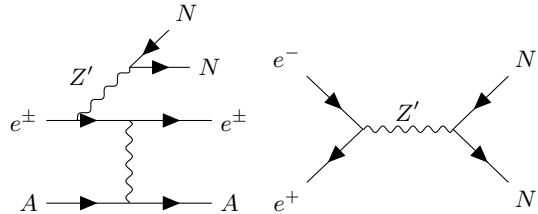

To calculate the number of $Z^\prime$ bremsstrahlung off of $N_{e^\pm}^{ij}$ electrons/positrons the $(i,j)$-th $(E_{e^\pm}^i, \theta_{e^\pm}^j )$ bin (where $\theta_{e^\pm}^j$ is the angle of the electron/positron with respect to the proton beam axis) as depicted in Fig.~\ref{fig:brem-res} left, we integrate over the GEANT4-simulated $e^\pm$ fluxes, convolving them with the differential cross section $\partial^2\sigma / (\partial E_{Z^\prime} \partial\Omega_{Z^\prime})$ for the process $e^\pm A \to e^\pm A Z^\prime$ off an atomic target $A$. The energy loss of the electrons and positrons in the material during particle transport must also be folded into the event rate calculation. The differential number flux of $Z^\prime$ can be expressed as  
\begin{align}
\label{eq:brem_flux}
    \dfrac{\partial^2 N_{Z^\prime}^{\text{brem}, ij}}{\partial E_{Z^\prime} \partial \Omega_{Z^\prime} } = \frac{N_A X_0}{A} \int_{m_e}^{E^i_{e^\pm}} &\int_0^{T} N_{e^\pm}^{ij}  I(t, E_{e^\pm}^{i}, E^\prime) \nonumber \\ &\times \dfrac{d^2\sigma(E^\prime)}{dE_{Z^\prime} d\Omega_{Z^\prime}} dt dE^\prime 
\end{align}
where $N_A$ is Avogadro's number, $X_0$ is the radiation length of the electrons/positrons in the dump material (g/cm$^2$), $A$ is the atomic weight (g/mol), and $N^{i}_{e^\pm}$ is the number of $e^\pm$ with energy $E_e^i$ and angle $\theta_e^j$ with respect to the beam axis. $I(t, E_i, E_f) = \frac{\theta(E_i - E_f)}{E_i \Gamma (4 t/3)} (\ln E_i/E_f)^{4t/3 - 1}$ is the energy loss smearing function, integrated over the dimensionless number of radiation lengths $t$ up to the radiation length of the target, $T$~\cite{PhysRevD.34.1326}. This expression gives us the differential rate of $Z^\prime$ over the differential angle $d\Omega_{Z^\prime} = d(\cos\theta_{Z^\prime}) d\phi_{Z^\prime}$ with respect to the $e^\pm$ direction $\theta_{e^\pm}^j$; we can then transform this angle to the lab frame in which the beam axis points along the $z$-direction via
\begin{equation}
    \theta_{Z^\prime}^{\rm lab} = \arccos (\cos\theta_{Z^\prime} \cos\theta_{e^\pm}^j + \cos\phi_{Z^\prime} \sin\theta_{Z^\prime} \sin\theta_{e^\pm}^j)\,.
    \label{eq:angle}
\end{equation}
Finally, we sum over the binned $e^\pm$ flux elements over $i,j$ bins.

\subsection{Resonant Production}
A $Z^\prime$ can be produced on-shell through the process $e^+ + e^-\to Z^\prime$ when $E^{\rm res}_{e^+}=m_{Z^\prime}^2/2m_e - m_e$ (Fig.~\ref{fig:brem-res}, right), which yields a mono-energetic $Z^\prime$ with energy $E_{Z^\prime} = m_{Z^\prime}^2 / 2 m_e$. In this case, the number of $Z^\prime$ generated from the $(i,j)$-th bin is given by
 \begin{equation}
 N_{Z^\prime}^{\text{res},ij}=\frac{ZX_0}{A} \int_{m_e}^{E^i_{e^+}} \int_{0}^{T} N_{e^+}^{ij}I(t, E^{i}_{e^+},E^\prime)\sigma_{\rm res} dt dE^\prime \,.
 \label{N_resonance_bin}
 \end{equation}
The prefactors and energy loss function are the same as those defined for electron/positron bremsstrahlung in \S~\ref{sec:ebrem}. $\sigma_{\rm res}$ is the cross section for resonant production and is given in \cite{Celentano:2020vtu};
\begin{equation}
\sigma_{\rm res}=\frac{\pi g_{B-L}^2}{2m_e}\delta\left(E^\prime -\frac{m^2_{Z^\prime}}{2m_e} + m_e\right)\,
\label{sigma_res}
\end{equation}
The delta function will then set $E^\prime = E_{e^+}^{\rm res}$ after integrating over $dE^\prime$. As before, we then sum over the $(i,j)$ bins of the positron flux to determine the total number of $Z^\prime$ produced at the monoenergetic resonant energy $E_{Z^\prime}$ and angle equal to the incoming positron angle $\theta_{e^+}^j$.

We note here that modeling the electron/positron energy loss for bremsstrahlung and resonant production using the track-length distribution function does not account for the random walk nature of the electromagnetic showers in a beam target. Recently, in ref.~\cite{Blinov:2024pza} showed using the \texttt{PETITE} package that more explicit modeling of the electromagnetic shower in the production of feebly-coupled vector particles can yield significantly different fluxes of the produced boson due to the random walk of the parent electrons/positrons deviating in their momentum direction in addition to losing energy as they transport through material. We can heuristically account for the broadening of the $Z^\prime$ angular distribution that would occur this way by limiting the number of track lengths integrated over in the above flux calculations. In the case of DUNE, we use a single value for $t_{\rm max}$, regardless of the original production point of positrons in the target. This average ($t_{\rm max}$) is calculated as the radiation lengths of the target downstream of the 50\% positron production point. The final result is $t_{\rm max} = 3.3$. In the case of the BNB dump, considering that secondary particles are propagating through the relatively thick beam dump, we take $t_{\rm max} = 5$. In each case, we find an agreement with the \texttt{PETITE}-derived fluxes to within 20\% difference.

\begin{figure}[ht]
  \includegraphics[width=0.5\textwidth]{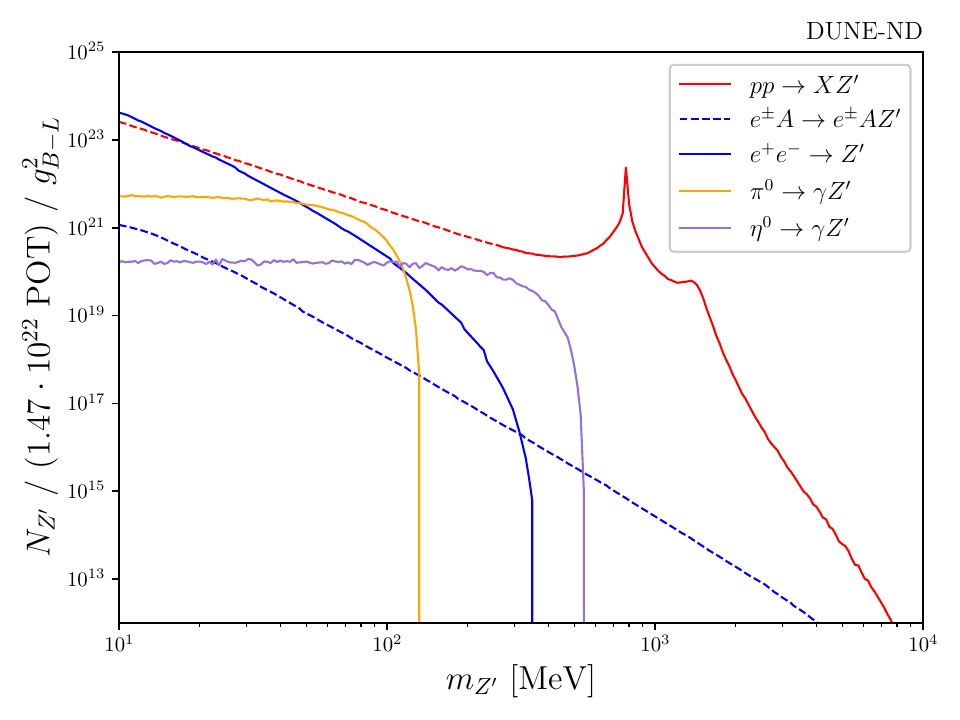}
 \caption{Number of $Z^\prime$ pointed within the solid angle of the DUNE near detector as a function of $m_{Z^\prime}$ for the different production channels, assuming $g_{B-L}=1$ and $1.47\times 10^{22}$ POT. In the case of proton bremsstrahlung, we do not use this channel below $m_{Z^\prime} = 250$ MeV as per~\cite{Foroughi-Abari:2021zbm}, but show the behavior of the event rate (dashed red) here for the interest of the reader.}
  \label{fig:dune_rates_by_Zp_mass}
\end{figure}

We show the rates of $Z^\prime$ that are produced and have 3-momentum that points within the solid angle of the DUNE near detector in Fig.~\ref{fig:dune_rates_by_Zp_mass} for resonant production, electron/positron bremsstrahlung, neutral meson decay ($\pi^0, \eta$), and proton bremsstrahlung as a function of the $Z^\prime$ mass. We find that the electron/positron channels contribute mainly to the low mass limit, while proton bremsstrahlung dominates the production for 250 MeV - 1 GeV masses. The extrapolated cross section for proton bremsstrahlung does grow at lower masses, shown by the dashed line in Fig.~\ref{fig:dune_rates_by_Zp_mass}, but we only consider this process above $m_{Z^\prime} > 250$ MeV, below which the appropriate behavior becomes theoretically uncertain.

\section{HNL decays in the detector}
\label{sec:detector}

In the parameter space we are interested in ($m_{Z^\prime}>20$ MeV and $g_{B-L}>10^{-7}$), the $Z^\prime$ decays promptly to leptons and HNL, and we, therefore, neglect its propagation distance before decay. Moreover, the $Z^\prime$ decays isotropically in the center of mass frame (cm). We define $\theta_{\rm HNL}$ and $\phi_{\rm HNL}$ as the angle with respect to the $z$-axis and the azimuthal angle, respectively, either in the center of mass frame or in the laboratory frame (lab). We assume that the angular distribution of the $Z^\prime$ only depends on its propagation angle with respect to the $z$-axis, whereas it is uniform in the azimuthal angle. Under these assumptions, the number of visible decays the HNL does in the detector is given by
\begin{widetext}
\begin{equation}
N_{HNL}=\sum_{ij}\frac{1}{4\pi}\int_0^{2\pi}{\rm d}\phi_{\rm HNL}^{\rm cm}\int_{-1}^1{\rm d}\cos\theta_{\rm HNL}^{\rm cm}\,N^{c,ij}_{Z^\prime}\,
{\rm Br}_{Z^\prime\to {\rm HNL}}\,
\Theta[\theta^{\rm lab}_{\rm HNL}+\theta^{\rm det}_{\rm min}]\,
\Theta[-\theta^{\rm lab}_{\rm HNL}+\theta^{\rm det}_{\rm max}]\,
P_{{\rm HNL}}(E_{\rm HNL}^{\rm lab})
\label{number_of_HNL}
\end{equation}
\end{widetext}
where $N^{c,ij}_{Z^\prime}$ is the number of $Z^\prime$ in the $ij$-th bin coming from channel $c$, ${\rm Br}_{Z^\prime\to {\rm HNL}}$ is the branching ratio of the decay of the $Z^\prime$ to HNL, $\Theta[x]$ is the Heaviside step function and $P_{\rm HNL}(E_{HNL})$ is the decay probability of the HNL with energy $E_{HNL}^{\rm lab}$.  We report in Table \ref{Tab:Exp_details} the values of $\theta^{\rm det}_{\rm min}$ and $\theta^{\rm det}_{\rm max}$ as well as the other detector specifications and exposures used for DUNE-ND, SBND, MicroBooNE, and ICARUS.

\begin{figure}[hb!]
    \centering
    \includegraphics[width=0.5\textwidth]{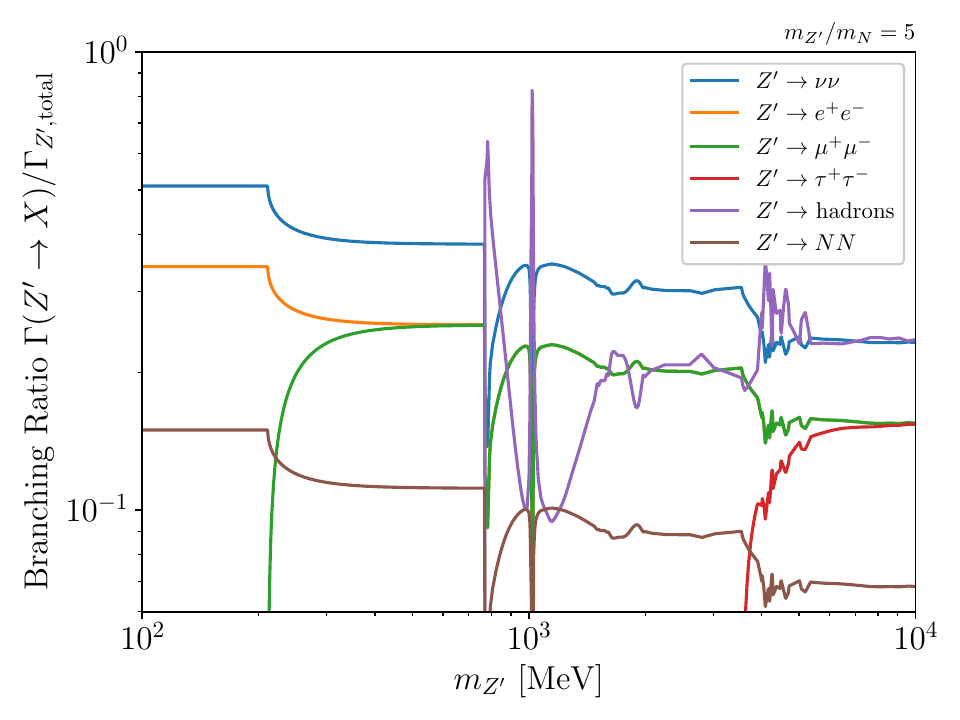}
    \caption{Branching ratios of the $B-L$ $Z^\prime$ to various final states including the majorana right-handed neutrino $N$. Here we take $m_{Z^\prime} = 5 \, m_N$.}
    \label{fig:zprime_br}
\end{figure}

\begin{figure*}[ht!]
    \centering
    \includegraphics[width=1.0\textwidth]{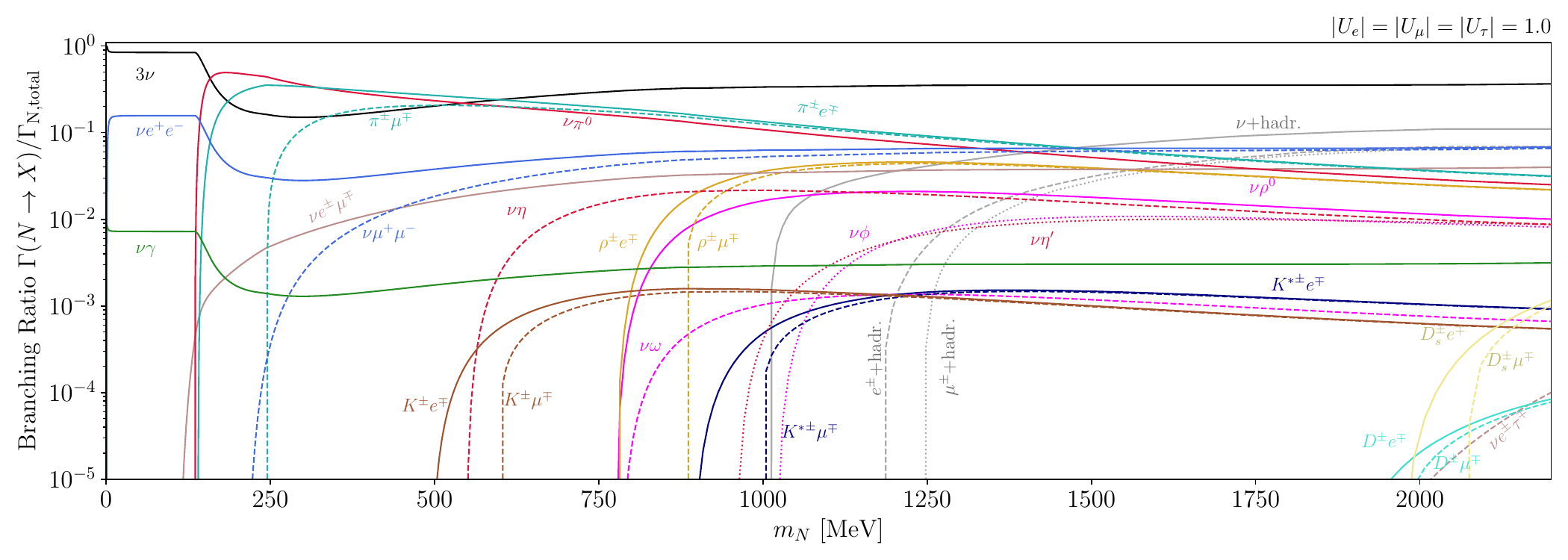}
    \caption{HNL branching fractions as a function of the HNL mass $m_N$. Final states are indicated on the plot, and all neutrino flavors have been summed over.}
    \label{fig:hnl_brs}
\end{figure*}

\begin{table*}[t]
\caption{\label{Tab:Exp_details} 
\footnotesize  Experimental details of DUNE and ICARUS used for evaluating Eq. \ref{number_of_HNL}. $L$ is the length of the detector, $d$ is the distance from the target or the beam dump to the detector, $w\times h$ is detector area perpendicular to the proton beam direction, and $\theta_{\rm det}^\text{max}$ is the approximate angle of the detector with respect to the proton beam axis determined by taking the radius of a circle whose area is the detector area. For DarkQuest, the tracking and calorimetry system is of variable dimension, so we use the area of the first tracking plane ($\sim 4$ m$^2$) to determine $\theta_\text{det}^\text{max}$, and use two benchmark exposure targets of $10^{18}$ and $10^{20}$ POT.}
\vspace*{0mm}
\centering
\begin{ruledtabular}
\begin{tabular}{c|cccccc}
Experiment & $d$ [m] & $L$ [m] & $w \times h$ [m$^2$] & $\theta^{\rm max}_{\rm det}$ [rad]  & POT\\
\hline
DUNE \cite{Berryman:2019dme}& 579 & 5 & $7\times 3$ & $4.3\times 10^{-3}$  &  $1.47\times 10^{22}$\\

SBND & 110 & 5 & $4 \times 4$ & $2.1 \times 10^{-2}$ & $6.6 \times 10^{20}$ \\
SBND Dump-mode & 60 & 5 & $4 \times 4$ & $3.8 \times 10^{-2}$ & $6.6 \times 10^{20}$ \\
MicroBooNE & 470 & 10.4 & $2.3\times 2.6$ & $2.9\times 10^{-3}$ &  $1.36 \times 10^{21}$ \\
ICARUS-BNB & 600 & 17.95 & $3.0 \times 3.16$ &  $2.8\times 10^{-3}$ &     $6.6 \times 10^{20}$ \\
DarkQuest & 4 & 14 & $\sim 4$ & $8 \times 10^{-2}$ & $10^{18}$ ($10^{20}$)
\end{tabular}
\end{ruledtabular}
\end{table*}

In the calculation of the total width of the decaying $Z^\prime$, we use the following partial decay width of the $Z^\prime$ to charged leptons 
\begin{align}
\Gamma(Z^\prime\to l^+l^-) &= \frac{g_{B-L}^2 m_{Z^\prime}}{12\pi} \sqrt{1 - 4\left(\frac{m_{l}}{m_{Z^\prime}}\right)^2} \nonumber \\
&\times \left[1 + 2 \left(\frac{m_{l}}{m_{Z^\prime}}\right)^2\right]\,,
\label{partial_decay_width_leptons}
\end{align}
where $l=e,\mu,\tau\,$. For the decay width of  $Z^\prime \to N N$, the decay width formula is calculated for the heavy $N$ final states;
\begin{align}
\Gamma(Z^\prime\to N N) &= \frac{g_{B-L}^2 m_{Z^\prime}}{24\pi} \sqrt{1 - 4\left(\frac{m_N}{m_{Z^\prime}}\right)^2} \nonumber \\
&\times \left[1 - \left(\frac{m_N}{m_{Z^\prime}}\right)^2\right]\,,
\label{partial_decay_width_HNL}
\end{align}
while the decay to the light neutrinos is simply
\begin{align}
\Gamma(Z^\prime\to \nu \nu) &= \frac{g_{B-L}^2 m_{Z^\prime}}{24\pi} 
\label{partial_decay_width_nus}
\end{align}
Lastly, the decays to hadrons are accessible for $m_{Z^\prime} > m_\rho = 770$ MeV. We use the $R$-ratio of hadronic to muonic final states in $e^+ e^-$ annihilation to parameterize the hadronic matrix elements as per refs.~\cite{Ilten:2018crw,Ezhela:2003pp,Bauer:2018onh};
\begin{equation}
    \Gamma(Z^\prime \to \text{hadrons}) = \frac{g_{B-L}^2}{e^2} \Gamma(Z^\prime \to \mu^+ \mu^-) R(m_{Z^\prime}^2)
\end{equation}
The branching ratios of the $Z^\prime$ to each of these final states are shown in Fig.~\ref{fig:zprime_br}.

Now we discuss the possibility of HNL decays in the detector. We consider all possible decay modes of the HNL mediated by its mixing angle $U_{\alpha}$ for $\alpha=e,\mu,\tau$. There are many modes which we do not list in their entirety here but are given in refs.~\cite{Coloma:2020lgy,Abdullahi:2023gdj,Berryman:2017twh}. Some dominant decay modes and the approximate range of HNL masses over which they are relevant ($\gtrsim 5$~\% branching ratio) are
\begin{align}
    N \to & \, \nu_\alpha \nu_\beta \bar{\nu}_\beta \, & m_N \gtrsim 1 \, {\rm eV} \nonumber \\
    N \to & \, \nu_\alpha e^+ e^- & m_N \in (1, 175)  \, {\rm MeV},\nonumber \\
    & &  \&  \, \, m_N \gtrsim 680 \, {\rm MeV} \nonumber \\
    N \to & \, \nu_\alpha \mu^+ \mu^- & m_N \gtrsim 900  \, {\rm MeV} \nonumber \\
    N \to & \, \pi^0 \nu_\alpha & m_N \in (150, 1700) \,{\rm MeV} \nonumber \\
    N \to & \, \pi^\pm e^\mp \, \, (\alpha=e) & m_N \in (150, 2000) \, {\rm MeV} \nonumber  \\
    N \to & \, \pi^\pm \mu^\mp \, \, (\alpha=\mu) & m_N \in (250, 2000) \, {\rm MeV} \nonumber  \\
    N \to & \, \nu_\beta e^\pm \mu^\mp \, \, (\alpha=e,\mu) &  m_N \gtrsim 2000  \, {\rm MeV} \nonumber \\
    N \to & \, \nu_\alpha + \, {\rm hadr.} &  m_N \gtrsim 1300  \, {\rm MeV} \nonumber \\
    N \to & \, e^\pm + \, {\rm hadr.} \, \, (\alpha=e)&  m_N \gtrsim 1600  \, {\rm MeV} \nonumber \\
    N \to & \, \mu^\pm + \, {\rm hadr.} \, \, (\alpha=\mu)&  m_N \gtrsim 1600  \, {\rm MeV} \nonumber
\label{eq:hnl_decay_final_states}
\end{align}
where ``hadr.''~refers to 3 or more mesons in the final state. Final states with flavor $\beta$ are associated with internal electroweak vertices and are summed over, and the relevant HNL mixing elements $U_\alpha$ that drive the decay are indicated wherever the $\alpha$-flavored neutrino does not appear in the final state, e.g., ``$\alpha = e,\mu$'' indicates that either $U_e$ or $U_\mu$ can drive the decay mode. We plot the branching ratios of each mode in Fig.~\ref{fig:hnl_brs} up to around 2.2 GeV in the HNL mass where the HNL becomes too short-lived and long baseline fixed target experiments begin to lose sensitivity.

The reader may also wonder if contributions to the HNL decay width can come from the $Z^\prime$ in addition to the mixing with electroweak diagrams. For the small gauge coupling we have adopted, $g_{B-L} \sim O(10^{-4})$, the contribution to the HNL branching ratios should go like $g_{B-L}^2 / m_{Z^\prime}^2 \ll g^2 / m_W^2$ for most of the parameter space we have considered. Since in the following analysis we consider $m_{Z^\prime} \gtrsim 100$ MeV, the contribution is limited in comparison with the electroweak factor of $g^2 / m_W^2 \simeq 6.5 \cdot 10^{-5}$ GeV$^{-2}$.

As for the case $m_{Z^\prime} < m_{HNL}$, this could be another interesting possibility that we do not consider here. In this case, the production mechanism could still arise from proton bremsstrahlung to a virtual $Z^\prime$ that pair produces $NN$, which would be phase space suppressed relative to the on-shell $Z^\prime$ we consider here, or via the usual charged meson decays with the mixing $U_\alpha$ driving the production rate. So, the light $m_{Z^\prime}$ limit would likely capture the same sensitivity as the $U_\alpha$-only sensitivity, however with the presence of some additional HNL decay channels that now get a less suppressed contribution from the $Z^\prime$ in addition to the exclusively electroweak and $U_\alpha$-driven decay modes (e.g. $N \to Z^\prime \nu_\alpha$). These new modes may change the lifetime slightly and offer new final states for the experiments to search for. Other signals could arise from long-lived $Z^\prime \to \ell^+ \ell^-$ decays, though these would be independent of the HNL decay signal and the HNL parameter space.

\section{Main results}\label{sec:results}

\begin{figure*}[ht!]
    \centering
\includegraphics[width=0.47\textwidth]{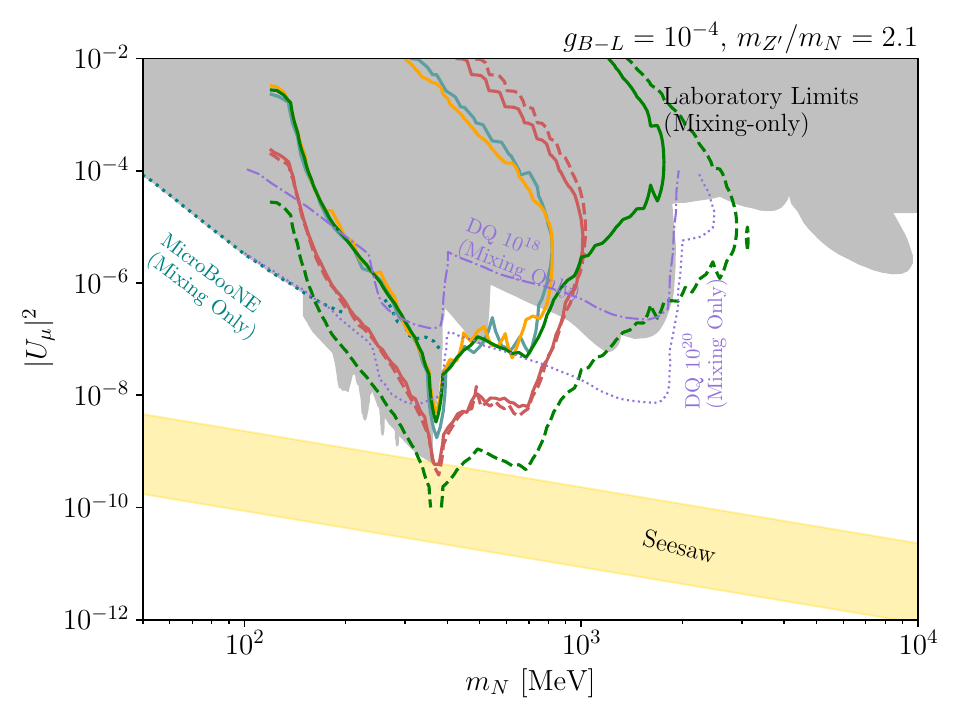}
\includegraphics[width=0.47\textwidth]{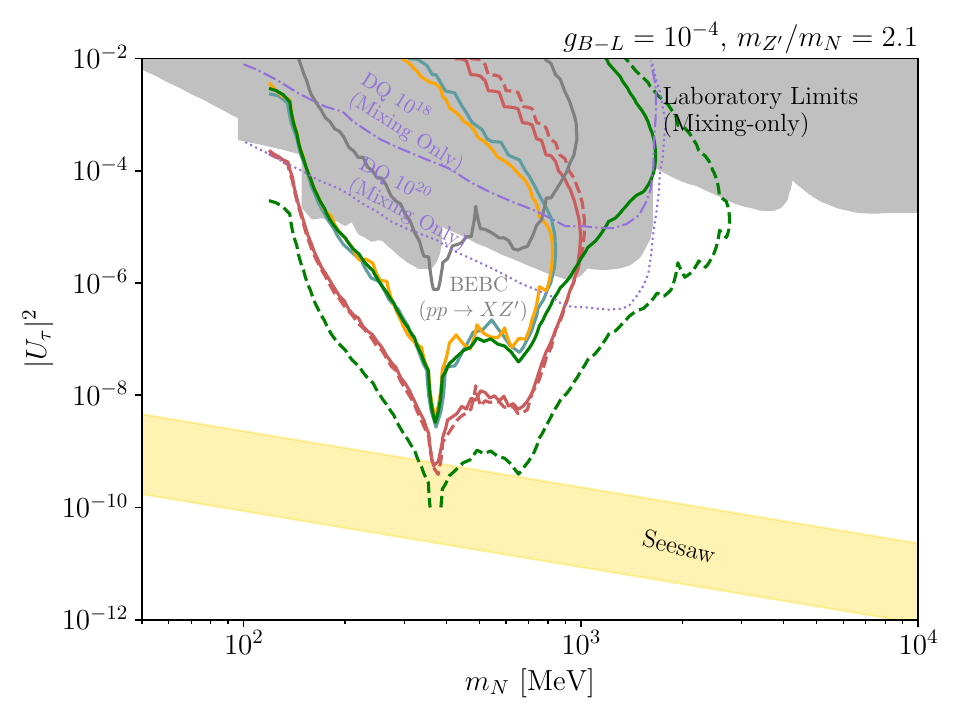}
\includegraphics[width=0.47\textwidth]{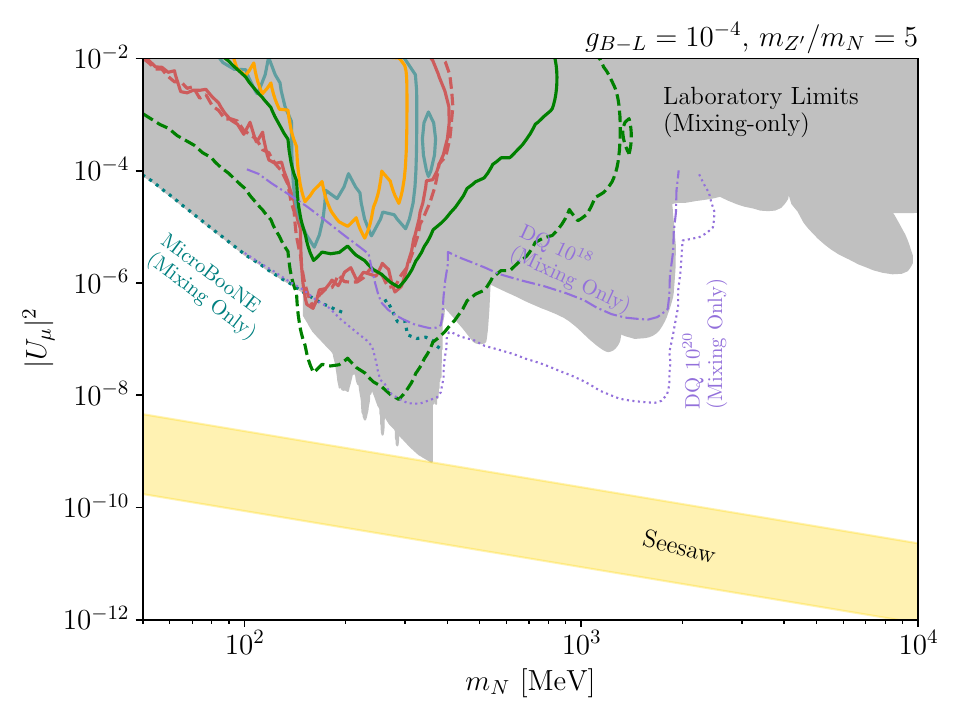}
\includegraphics[width=0.47\textwidth]{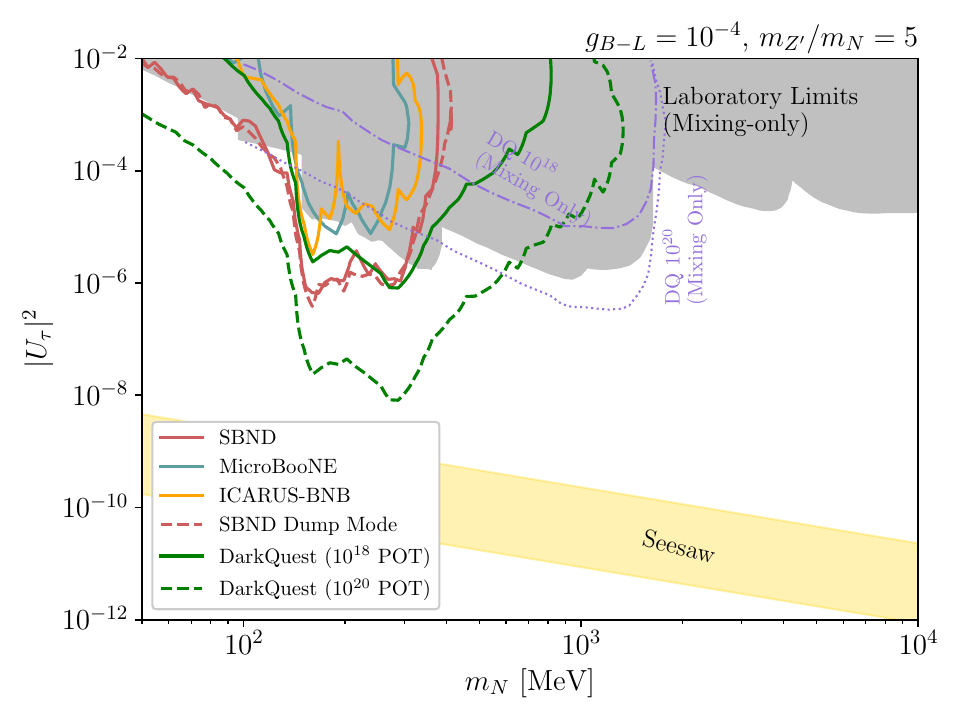}
    \caption{Background-free sensitivity contours at 90\% C.L. for SBND, MicroBooNE, ICARUS-BNB, and DarkQuest as a function of $m_N$ and $\left|U_\mu\right|^2$ (left column) and $\left|U_\tau\right|^2$ (right column). Top panels refer to $\frac{m_{Z^\prime}}{m_N}=2.1$ while bottom ones refer to $\frac{m_{Z^\prime}}{m_N}=5$. We fix the coupling of the $Z^\prime$ to $g=10^{-4}$. We also show the limit from MicroBooNE for HNL production driven by the $|U_\mu|$ mixing angle alone~\cite{MicroBooNE:2019izn,PhysRevD.104.055015}. For DarkQuest, we show the approved exposure benchmark of $10^{18}$ POT as well as the proposed, larger exposure of $10^{20}$ POT (green lines) in comparison with the forecasted limits using only the mixing (purple)~\cite{Batell:2020vqn}.}
    \label{fig:sbn_sensitivity}
\end{figure*}

We first show results for the SBN program (SBND, MicroBooNE, ICARUS-BNB) experiments, and for the DarkQuest experiment, in their sensitivity to HNL decays with dominant mixings to either muon ($U_\mu$) or tau ($U_\tau$) flavors for two mass ratio benchmarks $m_{Z^\prime}/m_N = 2.1$ and $m_{Z^\prime}/m_N = 5$ in Fig.~\ref{fig:sbn_sensitivity}, where the event rate is driven by all the visible modes of the decaying HNL, including both the leptonic and hadronic final states. In calculating this number, we have adopted a 20\% efficiency factor. This mimics the cuts one needs to apply to bring the background from neutrino interactions to a negligible level. In principle, the efficiency should depend on the decay channel and the mass of the HNL, but as shown in \cite{Coloma:2020lgy, Ballett:2019bgd}, using 20\% is still conservative. In addition, a timing measurement of the HNL signal has been demonstrated to significantly reduce backgrounds~\cite{SBND:2024vgn}. In the present work, we take this signal efficiency as an assumption, as it would be subject to change given a realistic dedicated background reduction analysis by the experiments. The sensitivity curves are then drawn from the iso-event contours corresponding to 3 events, or a 95\% confidence level of a zero Poisson background after the signal efficiency has been taken into account.

Here we have fixed the gauge coupling of the $Z^\prime$ to be $g_{B-L}=10^{-4}$, which is not yet excluded by laboratory bounds but well within potential reach of future $Z^\prime$ searches, e.g. LDMX and Belle-II~\cite{Nath:2021uqb}, though constraints from Texono, Borexino, and CHARM II do exclude this coupling for masses below $m_{Z^\prime} \lesssim 100$ MeV~\cite{Ilten:2018crw,Bauer:2018onh}. Therefore, we limit our considerations to heavier masses, or $m_N > 50$ MeV to be conservative given the mass ratio benchmarks mentioned above. For the SBN program experiments shown, which use the 8 GeV BNB proton beam, we do not find significant sensitivity from electron/positron bremsstrahlung, resonant production, or neutral meson decays; the primary sensitivity is due to $Z^\prime$ production from proton bremsstrahlung of the 8 GeV proton beam impinging on the beryllium target. We also indicate the region of the parameter space where light neutrino masses can be explained by the seesaw mechanism with the shaded band; the lower limit of this region is obtained by considering the standard seesaw formula 
\begin{equation}
\left|U_\alpha\right|^2=\frac{\sqrt{\Delta m^2_{21}}}{m_N}\,,
\label{eq:seesaw1}
\end{equation}
where $\Delta m^2_{21}=m^2_2-m^2_1\simeq 7.5\times 10^{-5}$ eV$^2$ is the mass difference measured by solar neutrino experiments and by KamLAND \cite{Super-Kamiokande:2016yck,SNO:2011hxd,KamLAND:2008dgz}. The upper limit is obtained by 

\begin{equation}
\left|U_\alpha\right|^2=\frac{0.23\,{\rm eV}}{m_N}\,,
\label{eq:seesaw2}
\end{equation}
where 0.23 eV is the current upper bound on the neutrino mass from Planck \cite{Planck:2018nkj}. 

We draw the existing constraints derived from solely considering the mixing $U_\alpha$ between the HNLs and the active neutrino flavors in gray~\cite{Bolton:2019pcu}. These come from a variety of different experiments, including beam target or beam dump experiments that looked for HNLs being both produced and detected through their mixings; i.e., their production through meson decays to HNLs in the beam target and their subsequent decays to the same final states that we consider in this work. These include limits from PIENU~\cite{Aguilar-Arevalo:2017vlf, Aguilar-Arevalo:2019owf}, PSI~\cite{Daum:1987bg}, E949~\cite{Artamonov:2014urb}, T2K~\cite{Abe:2019kgx}, NA62~\cite{NA62:2020mcv}, MicroBooNE~\cite{MicroBooNE:2019izn}, NuTeV~\cite{Vaitaitis:1999wq}, CHARM~\cite{Vilain:1994vg, Bergsma:1985is}, NOMAD~\cite{Astier:2001ck}, and BEBC~\cite{WA66:1985mfx, Barouki:2022bkt}, for example. One can also contrast the sensitivity from only considering the $B-L$ production channels (our curves) with the mixing-only sensitivity in this fashion, like the sensitivity derived by MicroBooNE~\cite{MicroBooNE:2019izn}, shown in Fig.~\ref{fig:sbn_sensitivity} by the teal dotted line.

We wish to make clear that the existing constraints shown in gray do not incorporate sensitivity to the gauge boson-driven production of HNLs, but in principle, a dedicated study for each experiment could be performed to recast their bounds with the physics of a $Z^\prime$ in mind. This would place the comparison with our forecasted iso-event contours within the same physics model, but at present, Fig.~\ref{fig:sbn_sensitivity} is meant to showcase how $Z^\prime$-driven sensitivity compares \textit{relative} to the mixing-only sensitivities. However, we argue that for many of the existing experiments the limits would not be enhanced significantly due to the presence of the $Z^\prime$. PSI, PIENU, and E949, for instance, all rely on the observation of $\pi$ or $K$ meson decays directly, and are more sensitive to the HNL mixing as opposed to e.g. $\pi^\pm \to Z^\prime \nu \ell^\pm$ decays, which are suppressed with couplings at the level of $10^{-4}$ through (if even kinematically allowed). Experiments like BEBC, PS191, T2K, CHARM, and NA62, however, constrain the heavier HNL masses and do involve powerful proton beam dumps that could potentially be sensitive to $Z^\prime \to NN$ production via $Z^\prime$ proton bremsstrahlung as we have considered. Out of these, since the T2K near detector (T2K-ND280) and CHARM are off-axis, while NA62 and PS191 make use of lower proton beam energies than BEBC, we estimate that BEBC would be the most promising of the existing experiments to gain $Z^\prime$-driven modifications to their limits. We performed a parameter scan using the BEBC detector and beam configurations~\cite{Barouki:2022bkt,WA66:1985mfx} and found the limit due to $Z^\prime$-produced HNLs in the $|U_\alpha|^2 - m_{HNL}$ plane is primarily relevant for the $U_\tau$ mixing at lower $m_{Z^\prime}/m_N$ mass ratios; see Fig.~\ref{fig:sbn_sensitivity}, upper right. The limit would modestly exclude parameter space beyond the mixing-only limits, and in line with the aforementioned argument, the other fixed target experiments mentioned are unlikely to see limits changed as significantly by the $Z^\prime$-driven HNL production. For the other benchmarks we found no additional sensitivity beyond the mixing-only limits from BEBC.

\begin{figure*}[ht!]
    \centering
\includegraphics[width=0.47\textwidth]{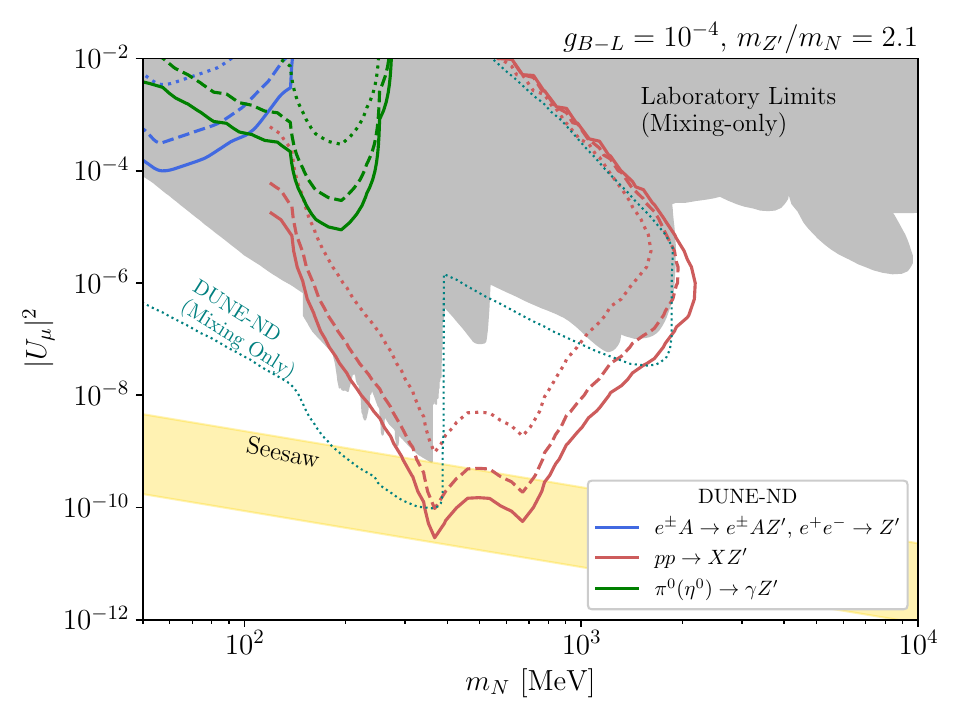}
\includegraphics[width=0.47\textwidth]{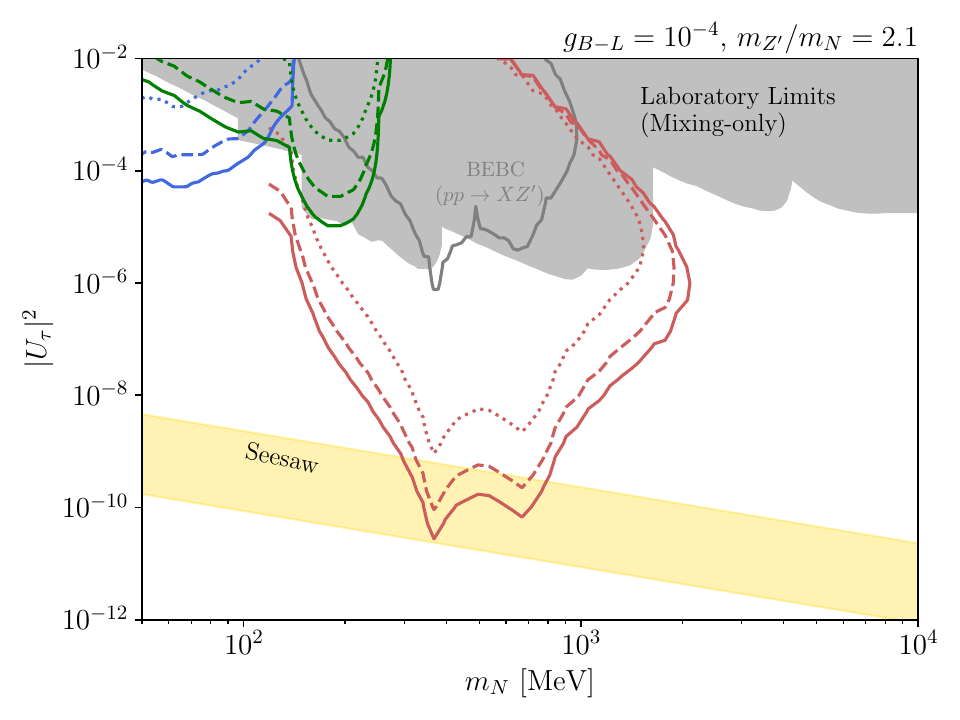}
\includegraphics[width=0.47\textwidth]{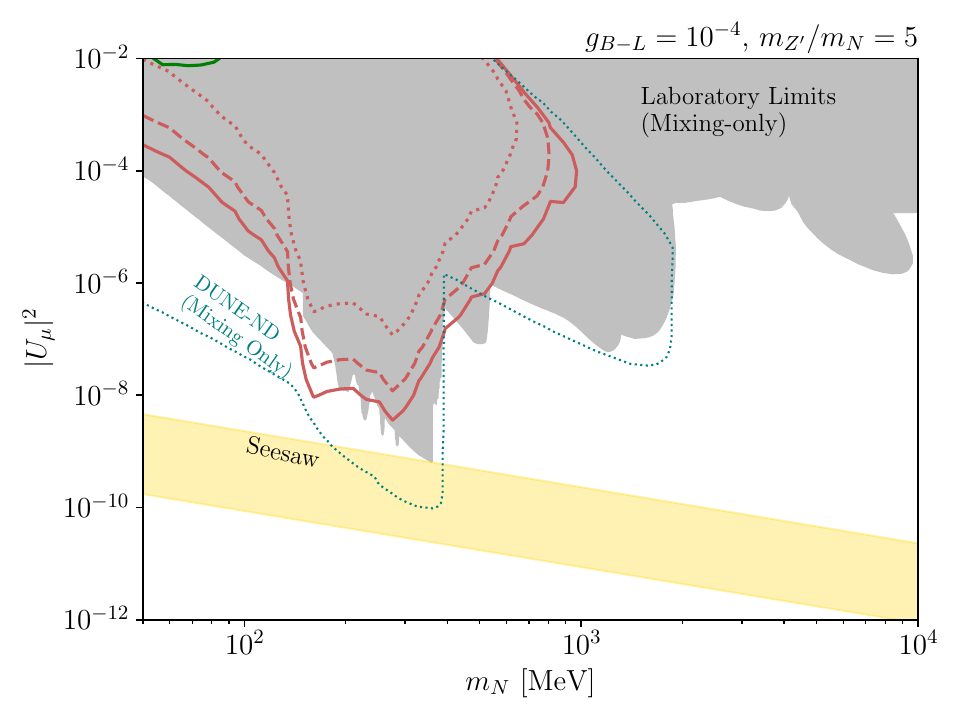}
\includegraphics[width=0.47\textwidth]{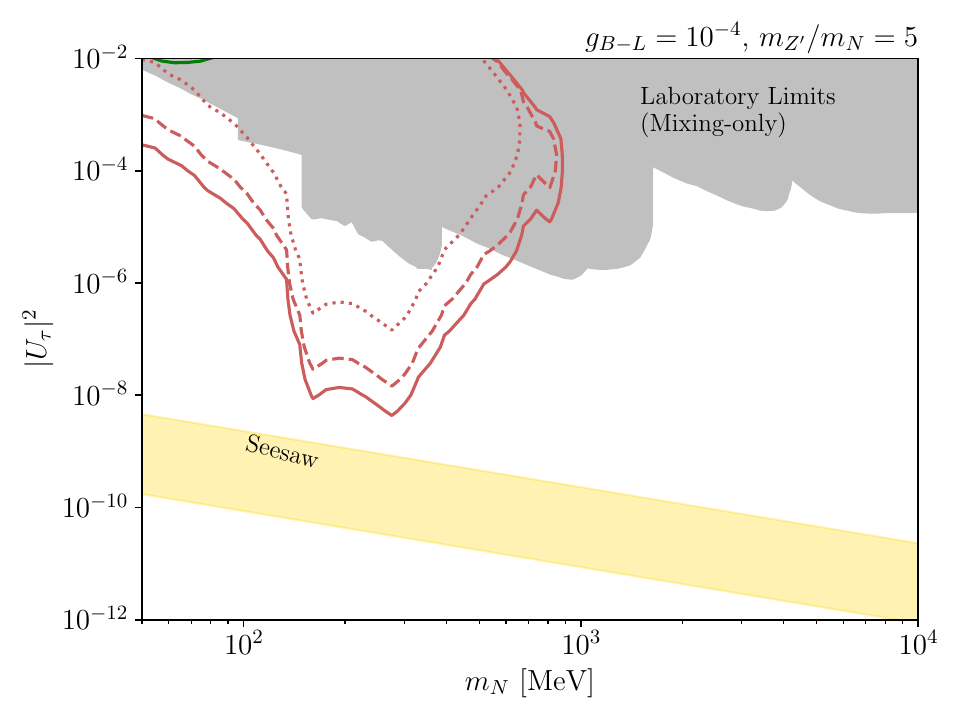}
    \caption{Number of particles from all visible decays in DUNE as a function of $m_N$ and $\left|U_\mu\right|^2$ (left column) and $\left|U_\tau\right|^2$ (right column). Top panels refer to $\frac{m_{Z^\prime}}{m_N}=2.1$ while bottom ones refer to $\frac{m_{Z^\prime}}{m_N}=5$. We fix the coupling of the $Z^\prime$ to $g=10^{-4}$. The solid, dashed, and dotted lines correspond to the iso-event contours for 3, 10, and 100 events, respectively, separated by production channel (neutral meson decays, electron/positron bremsstrahlung and resonant production, and proton bremsstrahlung). We also show the forecasted DUNE-ND sensitivity to HNLs produced only from the $|U_\mu|$ mixing-angle driven channels (teal dotted line)~\cite{Krasnov:2019kdc, Ballett:2019bgd}.}
    \label{fig:dune-nd_sensitivity}
\end{figure*}

Additionally, we wish to motivate the possibility of running SBND in a beam-dump or target-less mode, whereby the BNB target would be removed, and the magnetic horns turned off such that the proton beam impinges directly onto the steel beam dump. This configuration would be similar in spirit to the beam-dump run conducted at MiniBooNE~\cite{MiniBooNEDM:2018cxm} and the target-less configuration proposed for DUNE-ND in ref.~\cite{Brdar:2022vum} for their advantage in low neutrino backgrounds for BSM searches. The advantage of running in this mode lies in the much shorter distance between the HNL production site and the SBND detector, now foreshortened to only $\sim$50 meters. However, the beam power and exposure are expected to be limited in this configuration. As a realistic estimate, we take an exposure of $2\times 10^{20}$ POT and otherwise the same detector specifications. The resulting sensitivity for the SBND beam-dump mode run is shown by the dashed line in Fig.~\ref{fig:sbn_sensitivity}; we see that for only a third of the POT, the reach is roughly equivalent to the standard exposure in target mode.

Lastly, we show the DarkQuest iso-event contour in green in Fig.~\ref{fig:sbn_sensitivity} for benchmark exposures of $10^{18}$ POT and the proposed upgrade or extended run of $10^{20}$ POT (depending on the duty and plans for the Fermilab main injector beam). Note that this predicted event rate is again based on measuring all visible decay channels, both leptonic and hadronic, and we do not assume drastically different signal efficiencies between the two detection channels due to the detector capability. We find that the HNL production can also be enhanced from $B-L$ gauge boson production through proton bremsstrahlung compared to the electroweak production from neutrino mixing alone~\cite{Batell:2020vqn}. The $10^{20}$ POT benchmark, if it can be achieved, is especially attractive for its sensitivity to larger mass HNLs which is comparable to the DUNE-ND reach as we see next in Fig.~\ref{fig:dune-nd_sensitivity}.

We then show results for DUNE-ND sensitivity to HNL decays with dominant mixings to either muon or tau flavors in Fig.~\ref{fig:dune-nd_sensitivity}. Again, we show the iso-event contours for the same two mass ratio benchmarks $m_{Z^\prime}/m_N = 2.1$ and $m_{Z^\prime}/m_N = 5$, this time broken up into event rate levels of 3, 10, and 100 events observed over the integrated exposure. We again adopt a 20\% efficiency factor as we did for the SBN experiments. This time, the high-intensity beam and larger energies give us sensitivity to the $B-L$ driven production in both proton and $e^\pm$ bremsstrahlung. One can also consider the mixing-only sensitivity in this fashion for the DUNE-ND, which has been studied in refs.~\cite{Krasnov:2019kdc,Ballett:2019bgd}, shown in Fig.~\ref{fig:dune-nd_sensitivity} by the teal dotted line. 

We find that for smaller mass ratios between the $U(1)_{B-L}$ $Z^\prime$ mass and the HNL mass, the proton bremsstrahlung production channel at $g_{B-L} = 10^{-4}$ is strong enough to probe the seesaw band. This sensitivity is roughly independent of the dominant mixing flavor, $|U_\mu|$ or $|U_\tau|$, but is more sensitive to different gauge coupling strengths. For example, if the $B-L$ gauge coupling is reduced by a factor of 1/2, the sensitive C.L. on $|U_\alpha|^2$ would increase by 4 since the event rate is roughly proportional to $g_{B-L}^2 |U_\alpha|^2$ in the limit of long HNL lifetimes. Therefore we expect sensitivity to the seesaw band for gauge couplings bigger than roughly $5 \times 10^{-5}$ before the event rates would be too suppressed. In comparison with the mixing-produced HNL scenario projected for DUNE-ND in teal, we note that the $Z^\prime$-driven production of HNLs gives us sensitivity to larger HNL masses due to the kinematic reach of proton bremsstrahlung in addition to a completely new reach to $U_\tau$ flavor mixings. The existing constraints in the $U_\tau$-$m_N$ parameter space are somewhat weaker than for $U_\mu$ and $U_e$ such that the $Z^\prime$-driven production has a much stronger relative reach. \\

\section{Conclusions}
\label{sec:conclude}
While testing seesaw neutrino mass models has long been out of reach of modern experiments, many of the next generation of accelerator and beam target experiments designed to test neutrino physics and BSM physics are coming into position to test this parameter space for the first time. In this study, we point out that sensitivity to seesaw mass models, as well as to the greater parameter space of sterile neutrinos and generic HNLs, could be enhanced under the presence of extra gauge forces mediating HNL production in proton beam target environments.

By simulating the production of gauge bosons from proton bremsstrahlung, electron/positron bremsstrahlung and annihilation, and neutral meson decays, we have computed the resulting enhanced HNL flux from gauge boson decays. The flux is particularly enhanced from bremsstrahlung of the primary proton beams, a channel that sets gauged $B-L$ apart from leptophillic models, for example~\cite{Capozzi:2021nmp}. If the $B-L$ gauge coupling is in the neighborhood of the existing bounds, $10^{-4} - 10^{-5}$ and the gauge boson mass relative to the HNL mass is not too large, one could expect a significant enhancement to the expected HNL flux relative to the rates from standard electroweak production channels via the mixings $U_\alpha$. For a large enough coupling, these production rates are high enough to probe the parameter space of realistic seesaw models, while direct searches for $Z^\prime$ that couple to baryons and leptons are complementarily motivated and are well within future experimental reach, e.g. at Belle-II, LHCb, NA64$\mu$, and LDMX~\cite{Nath:2021uqb, Bauer:2018onh}. Sensitivity in this parameter space may be possible for experiments with high-intensity fluxes, as we have shown for DUNE, DarkQuest, and SBND. \\

\section*{Acknowledgements}
We thank Kevin J. Kelly for the useful feedback and discussions. We also thank Ekaterina Kriukova and Dmitry Gorbunov for their helpful comments. AT acknowledges support in part by the US Department of
Energy (DOE) grant \#DE-SC0010143. IMS is supported by the US Department of Energy under the award numbers DE-SC0020250 and DE-SC0020262. The work of F.C. was partially supported by the research grant number 2022E2J4RK ``PANTHEON: Perspectives in Astroparticle and Neutrino THEory with Old and New messengers" under the program PRIN 2022 funded by the Italian Ministero dell'Universit\`a e della Ricerca (MUR) and by the European Union - Next Generation EU. The work of BD is supported in part by the U.S.~Department of Energy Grant DE-SC0010813.
The work of GG, WJ and JY is supported by the U.S. Department of Energy under Grant No. DE-SC0011686. The code used for this research is made publicly available through \gitlink~under CC-BY-NC-SA~\cite{GitHNL}.
\bibliographystyle{bibi}
\bibliography{main}



\end{document}